\documentclass[oneside,reqno,a4paper,12pt]{amsart}

\usepackage{latexsym}
\usepackage{euscript}
\usepackage{epsfig}

\date{\today} 
 \large\normalsize  
\newcommand{\be}{\begin{equation}}
\newcommand{\ee}{\end{equation}}

\def\ie{{\it i.e.}}
\def\LEP2{{LEPII}}

\def\npb#1#2#3{    {\it Nucl. Phys. }{\bf B #1} (19#2) #3}

\def\plb#1#2#3{    {\it Phys. Lett. }{\bf B #1} (19#2) #3}
\def\prd#1#2#3{    {\it Phys. Rev. }{\bf D #1} (19#2) #3}
\def\prep#1#2#3{   {\it Phys. Rep. }{\bf #1} (19#2) #3}

\begin{document}
\begin{titlepage}  
\begin{flushright}
BA-99-40
\end{flushright}
\vskip 1.5cm
\title[]{Supersymmetric CP violating Phases and the LSP relic\\ \vskip
0.2cm density and
detection rates}

\maketitle

\begin{center}
\textsc{Shaaban Khalil$^{1,2}$ and Qaisar Shafi$^1$} \\
\vspace*{2mm} \small{\textit{$^1$Bartol Research Institute,
University of Delaware Newark, DE 19716}} \\ \vspace*{2mm}
\small{\textit{$^2$Ain Shams University, Faculty of Science, Cairo
11566, Egypt}} \\
\end{center}
\vspace*{2cm}
\begin{center}
ABSTRACT
\end{center}
\vspace*{8mm}
\begin{quotation}
For varying values of $\tan\beta$, we study the effect of CP
violating phases from the soft supersymmetry breaking terms in
string-inspired models on the relic abundance and detection rates
of the lightest neutralino (LSP). We find that the phases have no
significant effect on the LSP relic abundance but can have a
substantial impact on the detection rates.
\end{quotation} 
\setcounter{page}{1}
\end{titlepage}    
\section{Introduction}

        CP violation is an important test for physics beyond
the standard model. In the standard model, only one CP violating
phase exists in the Kobayashi-Maskawa matrix. However, in the
supersymmetric standard model there are many complex parameters,
in addition  to Yukawa couplings, which lead to new sources of CP
violation. These include the mass coefficient $\mu$ of the
bilinear term involving the two Higgs doublets, the $SU(3)$,
$SU(2)$ and $U(1)$ gaugino masses $M_3$, $M_2$ and $M_1$, and the
parameters $A_f$ and $B$ which respectively are the coefficients
of the supersymmetric breaking trilinear and bilinear couplings. (
the subscript $f$ denotes the flavor index.) \vskip 0.3cm In
minimal supersymmetric standard model (MSSM), only two of these
phases are physical. Through appropriate field redefinitions we
end up with the phase of $\mu$ ($\phi_{\mu}$) and the phase of $A$
($\phi_A$) as the physical phases which cannot be rotated away.
The phase of $B$ is fixed by the condition that $B\mu$ is real. It
is known that, unless these phases are sufficiently small, their
contributions to the neutron electric dipole moment (EDM) are
larger than the experimental limit $1.1\times 10^{-25}$ e.cm.
Recently, the effect of these phases on the EDM of the neutron
was examined in a model with dilaton-dominated supersymmetry
(SUSY) breaking, taking into account the cancellation mechanism
between the different contributions. It was shown that for a wide
region of the parameter space the phase of $\mu$ is constrainted
to be of order $10^{-1}$, while the phase of $A$ is strongly
correlated with that of $\mu$ in order not to violate the bound
on the neutron EDM. \vskip 0.3cm The effect of SUSY CP violating
phases on the relic density of the LSP has been considered for
the MSSM case in Ref.~\cite{falk1}, and for the supersymmetric
standard model coupled to N=1 supergravity in Ref.~\cite{falk2}.
It was shown in Ref.~\cite{falk1} that the upper bound on the LSP
(from $\Omega h^2 \leq 0.25$) is relaxed from 250 GeV to 650 GeV.
The effect of CP violation on the direct detection rates of the
LSP in MSSM is also considered in Ref.~\cite{nath2,falk3}. We
argue that such a large upper bound on the LSP is not possible in
the model we consider here. We show that in case of the bino-like
LSP the chance of the CP phases to have a significant effect is
very small. The impact of the phases on the direct and indirect
detection rates is an important issue and we present some details
here. \vskip 0.3cm  The paper is organized as follows. In section
2 we discuss the effect of the CP phases on the LSP mass and
purity within the string inspired model considered in
Ref.~\cite{barr}. In section 3 we compute the relic abundance of
the LSP for low and intermediate values of $\tan \beta$. We find
that the CP phases have almost no effect on the LSP relic
density, so that the upper bound on the LSP mass obtained in
Ref.~\cite{shafi,report} remains unchanged. In section 4 we
discuss the large $\tan \beta$($\simeq m_t/m_b$) case and again
find that there is no significant effect of CP phases on the LSP
relic density. In section 5 we show that CP phases can have a
substantial effect on the LSP detection rates. Our conclusions
are given in section 6

\section{String inspired model}

        We will consider the string inspired model which has been recently
studied in Ref.~\cite{barr}. In this model, the dilaton $S$ and
overall modulus field $T$ both contribute to SUSY breaking. The
soft scalar masses $m_i$ and the gaugino masses $M_a$ are given
as~\cite{munoz1}
\begin{eqnarray}
m^2_i &=& m^2_{3/2}(1 + n_i \cos^2\theta),
\label{scalar}\\
M_a &=& \sqrt{3} m_{3/2} \sin\theta e^{- i \alpha_{S}},
\label{gaugino}
\end{eqnarray}
where $m_{3/2}$ is the gravitino mass, $n_i$ is the modular
weight of the chiral multiplet, and $\sin \theta$ defines the
ratio between the $F$-terms of $S$ and $T$, (For example, the
limit $\sin \theta \rightarrow 1$ corresponds to a
dilaton-dominant SUSY breaking). The phase $\alpha_S$ originates
from the $F$-term of $S$.

The $A$-terms can be written as
\begin{eqnarray}
A_{ijk} &=& - \sqrt{3} m_{3/2} \sin\theta e^{-i \alpha_S}
- m_{3/2} \cos\theta
(3 + n_i + n_j + n_k) e^{-i \alpha_T},
\label{trilinear}
\end{eqnarray}
where $n_i$, $n_j$ and $n_k$ are the modular weights of the fields
that are coupled by this $A$-term. One needs a correction term in
eq~(\ref{trilinear}) when the corresponding Yukawa coupling
depends on moduli fields. However, the $T$-dependent Yukawa
coupling includes a suppression factor~\cite{vafa}, and so we
ignore it. Finally, the phase $\alpha_T$ originates from the
$F$-term of $T$. \vskip 0.3cm The magnitude of the soft SUSY
breaking term $B \mu H_1 H_2$ depends on the way one generates a
`natural' $\mu$-term. Here we take $\mu$ and $B$ as free
parameters and we will fix them by requiring successful
electroweak (EW) symmetry breaking. \vskip 0.3cm As stated
earlier, the gaugino masses as well as $A$-terms and $B$-term
are, in general, complex. We have the freedom to rotate $M_a$ and
$A_{ijk}$ at the same time~\cite{dugan}. Here we use the basis in
which $M_a$ is real. Similarly, we can rotate the phase of $B$ so
that $B\mu$ itself is real. In other words, $\phi_B = -\phi_{\mu}
= {\rm arg}(BM^*)$. In this basis, $A$-terms contain a single
phase, $(\alpha_A \equiv \alpha_T - \alpha_S)$. \vskip 0.3cm As
shown in eqs.(\ref{scalar}-\ref{trilinear}), the values of the
soft SUSY breaking parameters at string scale depend on the
modular weights of the matter states. The modular weights of the
matter fields $n_i$ are normally negative integers. Following the
approach of Ref.~\cite{lust} the `natural' values of modular
weights for matter fields ( in case of $Z_N$ orbifolds) are
-1,-2,-3 and -4. It was shown in Ref.~\cite{kobayashi} that the
following modular weights for quark and lepton superfields is
favorable for EW breaking $$n_Q=n_U=n_{H_1}= -1 ,$$ and
$$n_D=n_L=n_E=n_{H_2}=-2 .$$  Under this assumption we have
\begin{equation}
A_t=A_b= - \sqrt{3} m_{3/2} \sin\theta + m_{3/2} \cos\theta e^{-i
\alpha_A},
\end{equation}
and
\begin{equation}
A_{\tau}= - \sqrt{3} m_{3/2} \sin\theta + 2\ m_{3/2} \cos\theta
e^{-i \alpha_A},
\end{equation}
\vskip 0.3cm Given the boundary conditions in
eqs.~(\ref{scalar}-\ref{trilinear}) at the compactification
scale, we determine the evolution of the couplings and the mass
parameters according to their one loop renormalization group
equation in order to estimate the mass spectrum of the SUSY
particles at the weak scale. The radiative EW symmetry breaking
imposes the following conditions on the renormalized quantities:
\begin{equation}
m^2_{H_1} + m^2_{H_2} + 2\mu^2 > 2 B \mu,
\end{equation}
\begin{equation}
(m^2_{H_1}+\mu^2)( m^2_{H_2} + \mu^2) < (B \mu)^2,
\end{equation}
\begin{equation}
\mu^2= \frac{m^2_{H_1} - m^2_{H_2}\tan^2\beta}{\tan^2\beta -1} - \frac{M_Z^2}{2},
\end{equation}
and
\begin{equation}
\sin 2\beta = \frac{- 2 B \mu}{m^2_{H_1} + m^2_{H_2} + 2\mu^2},
\end{equation}
where $\tan\beta=\langle H_2^0 \rangle/\langle H_1^0 \rangle$ is
the ratio of the two Higgs VEVs that gives masses to the up and
down type quarks, and $m_{H_1}^2$, $m_{H_2}^2$ are the two soft
Higgs square masses at the EW scale. Using the above equations we
can determine $\vert \mu \vert$ and $B$ in terms of $m_{3/2}$,
$\theta$ and $\alpha_A$. The phase of $\phi_{\mu}$ remains
undetermined. \vskip 0.3cm Since we are interested in
investigating the effect of the supersymmetric phases $\alpha_A$
and $\phi_{\mu}$ on the relic density of the LSP and its direct
and indirect detection rates, we first study the allowed regions
of these phases and later impose the constraints (derived in
Ref~\cite{barr}) from the experimental bounds on the electric
dipole moments. \vskip 0.3cm The neutralinos $\chi_i^0$,
$(i=1,2,3,4)$ are the physical (mass) superpositions of the
Higgsinos $\tilde{H}_1^0$, $\tilde{H}_2^0$ and the two neutral
gaugino $\tilde{B}^0$ (bino) and $\tilde{W}_3^0$ (wino). The
neutralino mass matrix is given by
\begin{equation}
\hspace{-0.4cm}{\small M_N = \left(\begin{array}{clcr}M_1&0 &
-M_Z\cos\beta\sin\theta_W &M_Z\sin\beta\sin\theta_W\\
 0&M_2&M_Z\cos\beta\cos\theta_W&-M_Z\sin\beta\cos\theta_W\\
-M_Z\cos\beta\sin\theta_W &
M_Z\cos\beta\cos\theta_W&0&\mu e^{\phi_{\mu}}\\
 M_Z\sin\beta\sin\theta_W&-M_Z\sin\beta\cos\theta_W&\mu e^{\phi_{\mu}}&0
\end{array}\right),}
\label{neutralino}
\end{equation}
where $M_1$ and $M_2$ now refer to `low energy' quantities whose asymptotic
values are given in equation (\ref{gaugino}).
The lightest eigenstates $\tilde{\chi}^0_1$ is a linear
combination of the original fields:
\begin{equation}
\tilde{\chi}^0_1 = N_{11}\tilde{B}+ N_{12}\tilde{W}^3+
N_{13}\tilde{H}_1^0 + N_{14}\tilde{H}_2^0,
\end{equation}
where the unitary matrix $N_{ij}$ relates the $\tilde{\chi}^0_i$
fields to the original ones. The entries of this matrix depend on
$m_{3/2}$, $\theta$ and $\phi_{\mu}$. The dependence of the
$\tilde{\chi}^0_1$ (LSP) mass on $\phi_{\mu}$ is shown in figure
1 for $m_{3/2} \simeq 100 GeV$, $\cos^2 \theta \simeq 1/2$ and
$\alpha_A \simeq \pi/2$. \vskip 0.3cm
\begin{figure}[h]
\psfig{figure=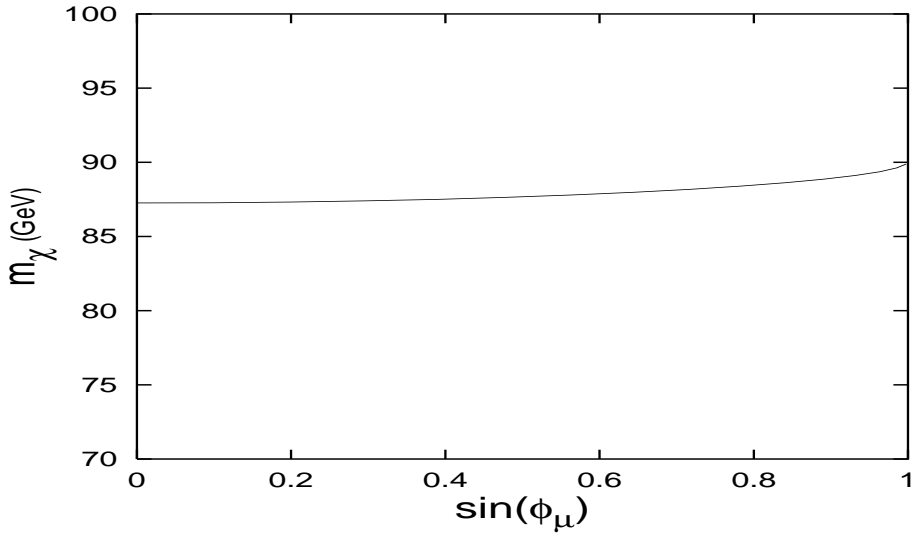,height=7cm,width=12cm} \caption{LSP mass as
a function of the phase $\phi_{\mu}$ .} \vskip 0.3cm
\end{figure}
A useful parameter for describing the neutralino composition is
the gaugino "purity" function
\begin{equation}
f_g= \vert N_{11}\vert^2 + \vert N_{12} \vert^2
\end{equation}
We plot this function versus $\phi_{\mu}$ in figure (2) which clearly shows
that the LSP is essentially a pure bino.
\begin{figure}[h]
\psfig{figure=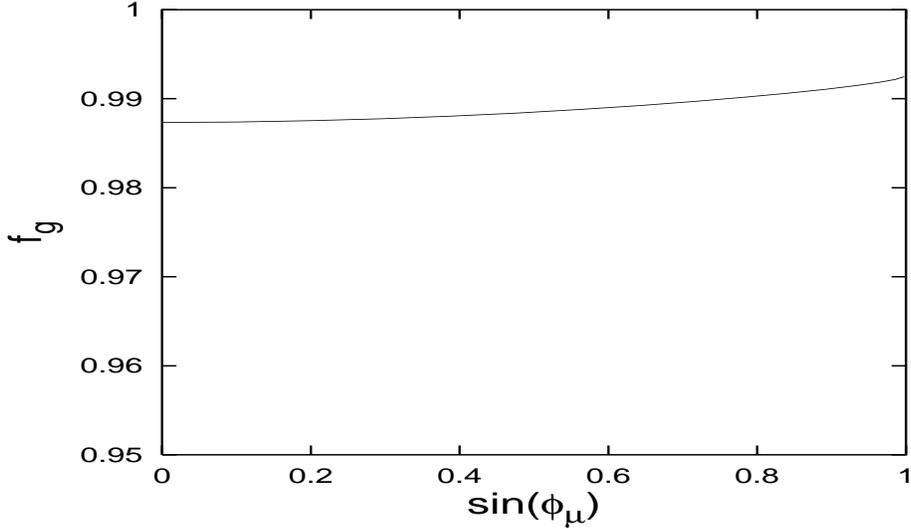,height=7cm,width=12cm} \caption{LSP gaugino
purity as a function of the phase $\phi_{\mu}$ .} \vskip 0.3cm
\end{figure}
These two figures show that the neutralino mass and composition
are only slightly dependent on the supersymmetric phase
$\phi_{\mu}$.

\section{Relic Abundance Calculation for low and\\ intermediate values of $\tan \beta$}
In this section we compute the relic density of the LSP  in the
case of low $\tan \beta$ (\ie\ $\tan \beta \simeq 3$), as well as
for intermediate $\tan \beta$ values (\ie\ $\tan \beta \simeq
15$). Using a standard method~\cite{report} in which we expand the
thermally averaged cross section $\langle \sigma_A v \rangle$ as
\begin{equation} \langle \sigma_A v \rangle = a + b v^2 + ...\ ,
\end{equation}
where $v$ is the relative velocity, $a$ is the s-wave contribution
at zero relative velocity and $b$ contains contributions from both
the s and p waves, the relic abundance is given by~\cite{report}.
\begin{equation}
\Omega_{\chi} h^2 = \frac{\rho_{\chi}}{\rho_c/h^2} = 2.82 \times 10^8
Y_{\infty} (m_{\chi}/GeV),
\end{equation}
where
\begin{equation}
Y_{\infty}^{-1}=0.264\ g_*^{1/2}\ M_P\
m_{\chi}\ (\frac{a}{x_F}+\frac{3b}{x_F^2}),
\end{equation}
$h$ is the well known Hubble parameter, $ 0.4 \leq h \leq 0.8$, and $
\rho_c \sim 2 \times 10^{-29} h^2$ is the critical density of the
universe. The freeze-out temperature is given by
\begin{equation}
x_F= \ln \frac{0.0764 M_P ( a+ 6 b/ x_F) c ( 2+c) m_{\chi}}{\sqrt{g_* x_F}}
\end{equation}
Here $ x_F=m_{\chi}/T_F$, $ M_P= 1.22 \times 10^{19} $ GeV is the
Planck mass, and $ g_*$ ($ 8 \leq \sqrt{g_*} \leq 10$) is the
effective number of relativistic degrees of freedom at $T_F$.
Also $c=1/2$ as explained in Ref~\cite{nojiri}. \vskip 0.3cm
Given that the LSP is bino-like, the annihilation is
predominantly into leptons, with the other channels either closed
or suppressed. It is worth noting that the squark exchanges are
suppressed due to their large masses, as figure (3) shows. \vskip
0.3cm
\begin{figure}[h]
\psfig{figure=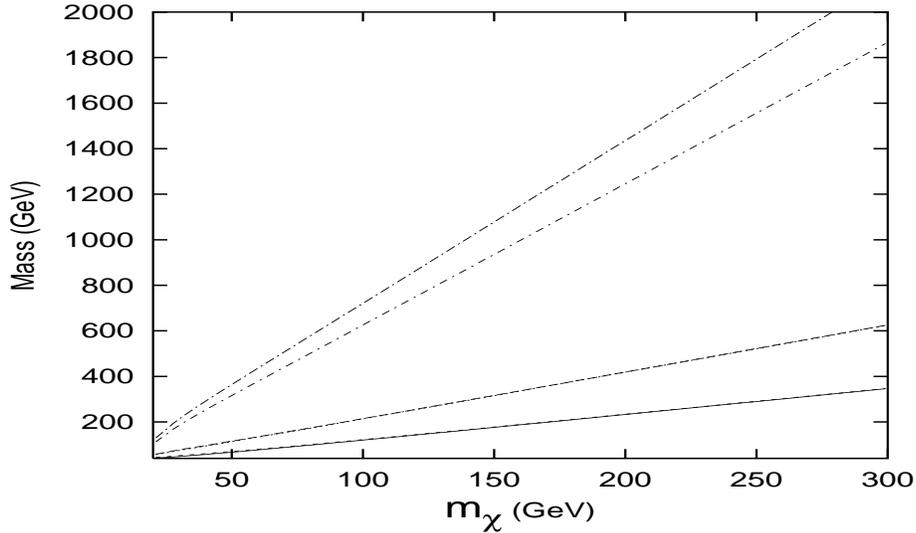,height=7cm,width=12cm} \caption{The squarks
(dashed-dotted) and sleptons (solid line) masses versus the  LSP
mass, for $\cos^2\theta \simeq 1/2$ and $\alpha_A \simeq \pi/2$ .
}\vskip 0.3cm
\end{figure}
The annihilation process is dominated by the exchange of the right
slepton. In fact, the masses of the right slepton are essentially
independent of $\alpha_A$ and $\phi_{\mu}$, unless there is a
significant amount of slepton mixing. Here, the off-diagonal
element of the matrices are much smaller than the diagonal
elements $M_{l_L}^2$ and $M_{l_R}^2$. Furthermore, since the LSP
is essentially a bino, it only slightly depends on the phase of
$\mu$ as figure (1) confirms. Therefore, we find that the
constraint on the relic density: $0.1 \leq \Omega_{{\rm LSP}}
\leq 0.9$, with $0.4 \leq h \leq 0.8$ leads, as figure (4) shows,
to the previously known upper bound on the LSP mass found in the
case of vanishing SUSY phases~\cite{shafi,falk2}, namely,
$m_{\chi} \leq 250$ GeV. \vskip 0.3cm
\begin{figure}[h]
\psfig{figure=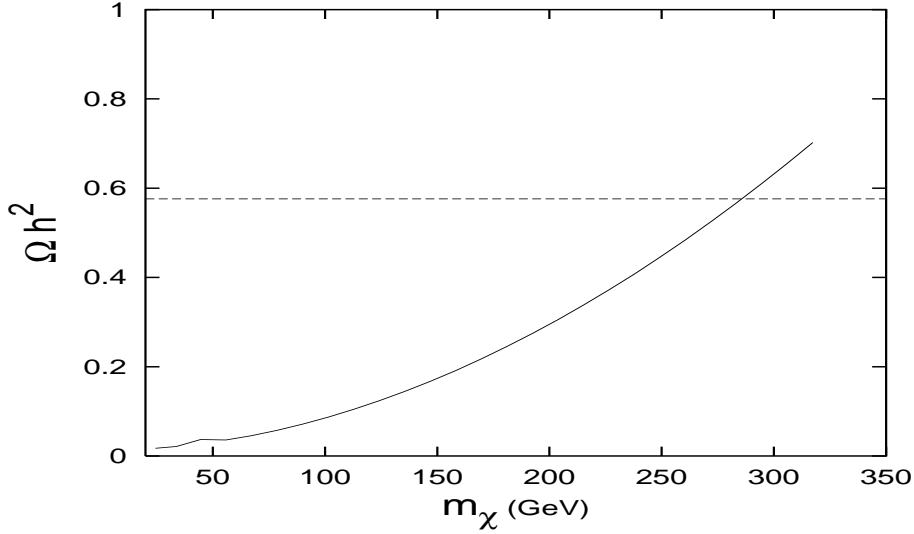,height=7cm,width=12cm} \caption{The relic
abundance of the LSP versus its mass.} \vskip 0.3cm
\end{figure}
This result is different from the one discussed in
Ref.~\cite{falk1} where it was claimed that the CP violating
phases  have a significant effect in that the cosmological upper
bound on the bino mass is increased from 250 GeV to 650 GeV. This
enhancement can be traced to the assumptions proposed in that
model, namely that all the scalar masses are equal and of order
$M_W$ at the weak scale. Also, the sfermion mxings were assumed
to be large (with $\mu \sim TeV$). It turns out that such
assumptions may lead to unacceptable charge and color breaking, as
explained in Ref.~\cite{falk2}. Moreover, they cannot be
motivated  from supergravity or superstring models. \vskip 0.3cm
We have also considered the relic density for intermediate $\tan
\beta$ (\ie\ $\tan \beta \simeq 15$). We find that there is no
significant difference between this and the case of low $\tan
\beta$. The upper bound on the LSP mass is still of order 250 GeV.

\section{SUSY CP phases with Large $\tan \beta$}
We now extend our study to the case when $\tan \beta$ is large
($\simeq 50$). In a large class of supersymmetric models with
flavor $U(1)$ symmetry, $\tan \beta \sim (\frac{m_t}{m_b}).
\epsilon^{n}$, where $\epsilon \simeq 0.2$ ($\simeq$ Cabibbo
angle) is a `small' expansion parameter, and $n=0,1,2$. (See for
instance \cite{zt} and reference therein). \vskip 0.3cm For $\tan
\beta \simeq \frac{m_t}{m_b}$, it is known that the
phenomenological aspects of these models are very different
compared with the small $\tan \beta$ case. In particular,
radiative EW symmetry breaking is an important non-trivial issue.
Non-universality, such as $m_{H_1}^2
> m_{H_2}^2$ at the Planck scale, is favored for a successful EW
breaking with large $\tan \beta$. Further, non-universality of the
squark and slepton masses can affect symmetry breaking as well as
other phenomenological aspects. We have adopted this
non-universality in our choice for the modular weights in section
2. \vskip 0.3cm In the large $\tan \beta$ case the Higgs potential
has two characteristic features. It follows from the minimization
conditions that
\begin{equation}
m_2^2 \simeq -\frac{M_Z^2}{2},
\label{cons1}
\end{equation}
\begin{equation}
m_3^2 \simeq \frac{M_A^2}{\tan^2 \beta}\sim 0,
\label{cons2}
\end{equation}
with
\begin{equation}
M_A^2 \simeq m_1^2 +m_2^2 >0.
\label{cons3}
\end{equation}
Here, $m_i^2 = m_{H_i}^2 + \mu, i=1,2$ and $m_3^2 = B \mu$. A
combination of eqs.(\ref{cons1}) and (\ref{cons3}) gives the
following constraint on the low energy parameters
\begin{equation}
m_1^2 -m_2^2 > M_Z^2 ,
\label{cons4}
\end{equation}
i.e $m_{H_1}^2 -m_{H_2}^2 > M_Z^2$. In order to have electroweak
breaking in the large $\tan \beta$ case, the difference between
the masses of the two Higgs fields should satisfy the above
inequality. \vskip 0.3cm
 In our model we find that this inequality
is indeed satisfied, and the EW symmetry is broken at the weak
scale. Also, one of the stau leptons ($\tilde{\tau}_R$) has a
`small' mass of order O(100) GeV, and happens to be the lightest
slepton. It therefore dominates the LSP annihilation process.
This relaxes the upper bound on the LSP mass from 250 to 300 GeV.
Thus, even in the of large $\tan \beta$ case the effect of the
supersymmetric phases are relatively small as figure (5) shows.
This essentially follows because the diagonal elements of the
stau mass matrices, in this model are larger than the
off-diagonal ones, \ie, there is no large mixing, as well as from
the fact that the LSP is bino like. \vskip 0.3cm
\begin{figure}[h]
\psfig{figure=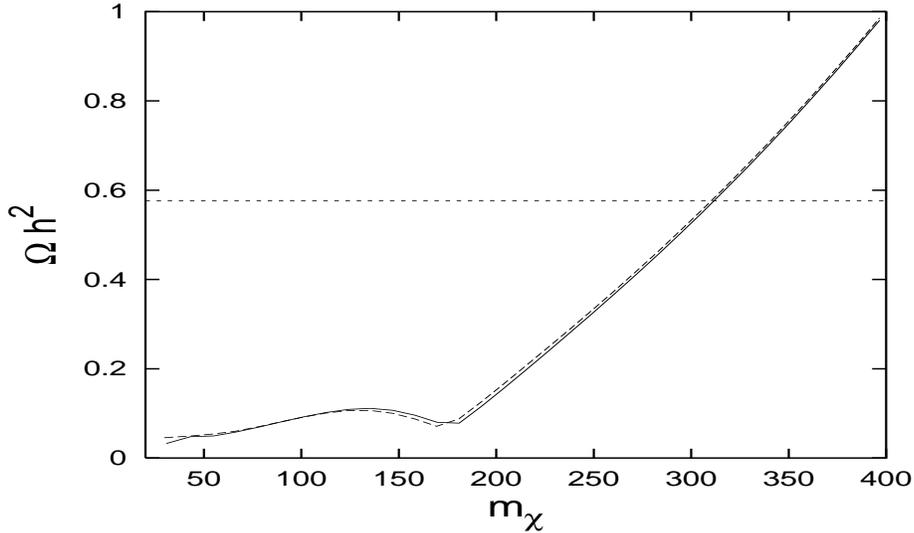,height=7cm,width=12cm} \caption{The relic
abundance with non-vanishing phases (solid line) and vanishing
phases (dashed line) versus the LSP mass in case of
$\tan\beta\simeq \frac{m_t}{m_b}$.}\vskip 0.3cm
\end{figure}

\section{ CP phases and detection rates of the LSP}

We have seen that the effects of CP violating phases on the
neutralino relic density are very small. In this section we
examine the effect of these phases on the event rates of relic
neutralinos scattering off nuclei in terrestrial detectors. The
direct detection experiments provide the most natural way of
searching for the neutralino dark matter. Any large CP violating
phases can affect the detection rate, as will see below. It is
interesting to note that the measured event rate  may shed light
on the value of the supersymmetric phases. \vskip 0.3cm The
differential detection rate is given by~\cite{report}
\begin{equation}
\frac{d R}{d Q} = \frac{\sigma \rho_{\chi}}{2 m_{\chi} m_r^2} F^2(Q)
\int_{v_{min}}^{\infty} \frac{ f_1(v)}{v} dv,
\label{rate}
\end{equation}
where $f_1(v)$ is the distribution of speeds relative to the detector.
The reduced mass is $m_r= \frac{m_{\chi} m_N}{m_{\chi}^2+ m_r^2}$, where $
m_N$ is the mass of the nucleus, $ v_{min}= (\frac{Q m_N}{2 m_r^2})^{1/2}$, $Q$
is the energy deposited in the detector, and $\rho_{\chi}$ is the
density of the neutralino near the Earth. $\sigma$ is the elastic-scattering cross
section of the LSP with a given nucleus. In
general $\sigma$ has two contributions: a spin-dependent contribution
arising from $Z^0$ and $\tilde{q}$ exchange diagrams, and a spin-independent
(scalar) contribution due to the Higgs and squark exchange diagrams. For
$^{76}Ge$ detector, where the total spin of $^{76}Ge$ is equal to zero,
we have contributions only from the scalar part.
\begin{equation}
\sigma = \frac{4 m_r^2}{\pi} [ Z f_p + (A-Z) f_n ]^2,
\end{equation}
where $Z$ is the nuclear charge, and $A-Z$ is the number of
neutrons. The expressions for $f_p$ and $f_n$, and their
dependence on the SUSY phases can be found in Ref.~\cite{nath2,
falk3}. The effect of the CP violating phases enter through the
neutralino eigenvector components $N_{ij}$, and also through the
matrices that diagonalize the squark mass matrices. Finally,
$F(Q)$ in (\ref{rate}) is the form factor. We use the standard
parameterization~\cite{engel}
\begin{equation}
F(Q) = \frac{3 j_1(q R_1) }{ q R_1} e^{-\frac{1}{2} q^2 s^2},
\end{equation}
where the momentum transfer $q^2= 2 m_N Q$, $R_1=(R^2 -5
s^2)^{1/2}$ with $R=1.2 fm A^{1/2}$, and A is the mass number of
$^{76}Ge$. $j_1 $ is the spherical Bessel function and $s \simeq 1
fm $. \vskip 0.3cm The ratio $R$ of the event rate with
non-vanishing CP violating phases to the event rate in the absence
of these phases is  presented in  figure (6). The solid curve
corresponds to the case $\alpha_A=0$, while the dashed one
corresponds to $\alpha_A=\pi/2$. \vskip 0.3cm
\begin{figure}[h]
\psfig{figure=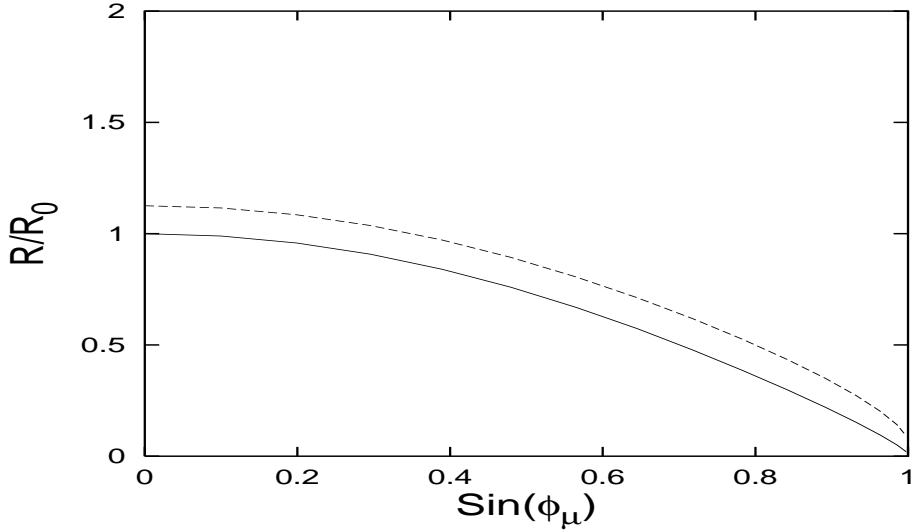,height=7cm,width=12cm} \caption{The ratio
of the direct detection rates as function of $\phi_{\mu}$.}
\end{figure}
\vskip 0.3cm From this figure, it is clear that, in the model we
are considering, the CP violating phases can significantly affect
the event rates of the direct detection of the LSP. The phase
$\phi_{\mu}$ reduces the value of $R$, while the phase $\alpha_A$
in the trilinear coupling increases it. However, as explained in
Ref.~\cite{barr}, $\phi_{\mu}$ is constrained from the electric
dipole moment experimental limit to be $\leq 10^{-1}$ \vskip
0.3cm For completeness, we also examine the effect of CP
violating phases on the indirect detection rates of the LSP in
the halo. The observation of energetic neutrinos from the
annihilation of the LSP that accumulate in the sun or in the
earth is a promising method for detecting them. The technique for
detecting such energetic neutrinos is through observation of
upward going muons produced by charged current interactions of the
neutrinos in the rock below the detector. The flux of such muons
from neutralino annihilation in the sun is given by
\begin{equation}
\Gamma = 2.9 \times 10^8 m^{-2} yr^{-1} \tanh^2(t/\tau) \rho_{\chi}^{0.3}
f(m_{\chi}) \zeta(m_{\chi}) (\frac{m_{\chi}^2}{GeV})^2
(\frac{f_P}{GeV^{-2}})^2 .
\end{equation}
The neutralino-mass dependence of the capture rates is described
by~\cite{report}
\begin{equation}
f(m_{\chi}) = \sum_i f_i \phi_i S_i(m_{\chi}) F_i(m_{\chi}) \frac{m_i^3
m_{\chi}}{(m_{\chi}+m_i)^2},
\end{equation}
where the quantities $\phi_i$ and $f_i$ describe the distribution
of element $i$ in the sun and they are listed in
Ref.~\cite{report}, the quantity
$S_i(m_{\chi})=S(\frac{m_{\chi}}{m_{N_i}})$ is the kinematics
suppression factor for the capture of neutralino of mass $
m_{\chi}$ from a nucleus of mass $m_{N_i}$~\cite{report}, and
$F_i(m_{\chi})$ is the form factor suppression for the capture of
a neutralino of mass $ m_{\chi}$ by a nucleus $i$. Finally, the
function $\zeta (m_{\chi})$ describes the energy spectrum from
neutralino annihilation for a given mass.\\ \vskip 0.3cm
\begin{figure}[h]
\psfig{figure=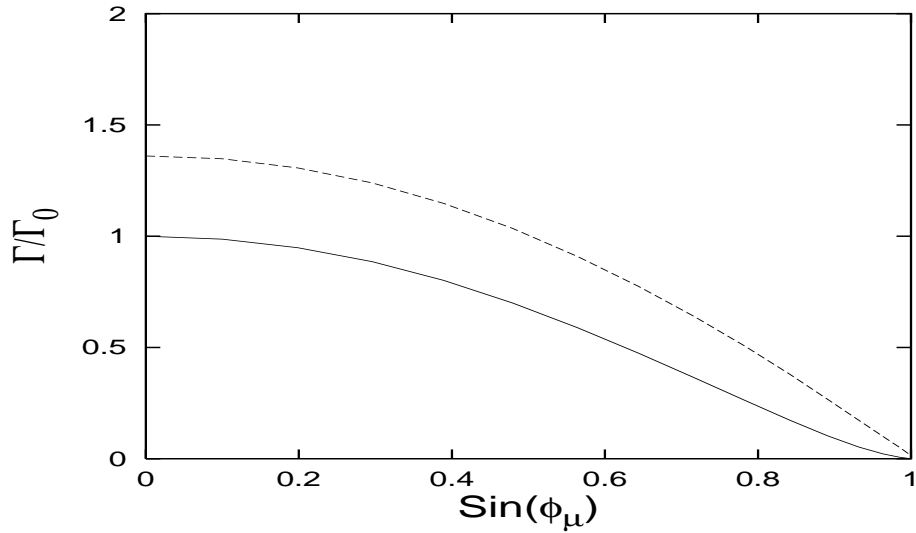,height=7cm,width=12cm} \caption{The ratio
of the indirect detection rates as function of $\phi_{\mu}$}
\end{figure}
\vskip 0.3cm
        In Figure (7) we present the ratio of the muon fluxes
resulting from captured neutralinos in the sun in the case of
non-vanishing CP violating phases to that of vanishing $\phi_A$,
for $\rho_{\chi}=0.3 GeV/cm^3$. We see that the predicted muon
flux increases as the phase of A-term is increased. \vskip 0.3cm
We can understand this important effect of CP violating phases on
the detection rate as follows. The phases affect the neutralino
eigenvector components $N_{ij}$ and the squark mass matrices.
Consequently, they have a significant effect on the neutralino
coupling to quarks. The spin independent contribution, as also
shown in Ref.~\cite{nath2} and~\cite{falk3}, is decreased by
increasing the phase of $\mu$, and goes in the other direction if
the phase of $A$-term is increased. This leads to the same
behaviour for the elastic scattering cross section, which
translates this dependence on the phases of $\mu$ and $A$ to the
detection rates as figures 6 and 7 confirm.

\section{Conclusions}
We have studied the impact of CP violating phases from soft SUSY
breaking terms in string-inspired models on the LSP, its purity
and its relic abundance density. For different values of $\tan
\beta$ (of order unity, intermediate and of order $m_t/m_b$), we
found that these phases have no significant effect on the LSP
relic density, so that the upper bound on the LSP mass is
essentially unchanged. We have examined the effect of these
phases on the direct and indirect detection rates. We found that
increasing the value of the phase $\phi_{\mu}$ leads to a decrease
the event rates, while the phase $\phi_A$ of the trilinear
coupling has the opposite effect.

\section*{Acknowledgments}
S.K. would like to acknowledge the support provided by the
Fulbright Commission and the hospitality of the Bartol Research
Institute. Q.S. is supported in part by DOE Grant No.\
DE-FG02-91ER40626, and the Nato contract number CRG-970149.

\providecommand{\bysame}{\leavevmode\hbox
to3em{\hrulefill}\thinspace}

\end{document}